\documentclass[fleqn,10pt]{wlscirep}

\usepackage[utf8]{inputenc}
\usepackage[T1]{fontenc}
\usepackage{graphicx}
\usepackage{dcolumn}
\usepackage{bm}
\usepackage{hyperref}
\usepackage{xcolor}

\usepackage{lipsum}

\newcommand\blfootnote[1]{%
  \begingroup
  \renewcommand\thefootnote{}\footnote{#1}%
  \addtocounter{footnote}{-1}%
  \endgroup
}

\def \bfso{Ba$_2$FeSi$_2$O$_7$}

\title{Understanding temperature-dependent SU($3$) spin dynamics in the $S=1$ antiferromagnet Ba$_2$FeSi$_2$O$_7$}

\author[1,*]{Seung-Hwan Do}
\author[1,2,*]{Hao Zhang}
\author[2]{David A. Dahlbom}
\author[3]{Travis J. Williams}
\author[3]{V. Ovidiu Garlea}
\author[3]{Tao Hong}
\author[4]{Tae-Hwan Jang}
\author[4,5]{Sang-Wook Cheong}
\author[4,6]{Jae-Hoon Park}
\author[7]{Kipton~Barros}
\author[2,$\dag$]{Cristian D. Batista}
\author[1,$\ddag$]{Andrew D. Christianson}

\affil[1]{Materials Science and Technology Division, Oak Ridge National Laboratory, Oak Ridge, Tennessee 37831, USA} 
\affil[2]{Department of Physics and Astronomy, University of Tennessee, Knoxville, Tennessee 37996, USA}
\affil[3]{Neutron Scattering Division, Oak Ridge National Laboratory, Oak Ridge, Tennessee 37831, USA}
\affil[4]{MPPHC-CPM, Max Planck POSTECH/Korea Research Initiative, Pohang 37673, Republic of Korea}
\affil[5]{Rutgers Center for Emergent Materials and Department of Physics and Astronomy, Rutgers University, Piscataway, New Jersey 08854, USA}
\affil[6]{Department of Physics, Pohang University of Science and Technology, 37673, Republic of Korea}
\affil[7]{Theoretical Division and CNLS, Los Alamos National Laboratory, Los Alamos, New Mexico 87545, USA}

\affil[*]{These authors contributed equally to this work}
\affil[$\dag$]{cbatist2@utk.edu}
\affil[$\ddag$]{christiansad@ornl.gov}

\begin{abstract}
Quantum magnets admit more than one classical limit and $N$-level systems with strong single-ion anisotropy are expected to be described by a classical approximation based on SU($N$) coherent states. 
Here we test this hypothesis by modeling finite temperature inelastic neutron scattering (INS) data of the effective spin-one antiferromagnet \bfso{}. The measured dynamic structure factor is calculated with a  generalized Landau-Lifshitz dynamics for SU($3$) spins. Unlike the traditional classical limit based on SU($2$) coherent states, the results obtained with classical SU($3$) spins are in good agreement with the measured temperature dependent spectrum. The SU($3$) approach developed here provides a general framework to understand the broad class of materials comprising weakly coupled antiferromagnetic dimers, trimers, or tetramers, and magnets with strong single-ion anisotropy.
\end{abstract}
\begin{document}

\flushbottom
\maketitle

\thispagestyle{empty}

\blfootnote{This manuscript has been authored by UT-Battelle, LLC under Contract No. DE-AC05-00OR22725 with the U.S. Department of Energy.  The United States Government retains and the publisher, by accepting the article for publication, acknowledges that the United States Government retains a non-exclusive, paid-up, irrevocable, world-wide license to publish or reproduce the published form of this manuscript, or allow others to do so, for United States Government purposes.  The Department of Energy will provide public access to these results of federally sponsored research in accordance with the DOE Public Access Plan (http://energy.gov/downloads/doe-public-access-plan).}

\section*{Introduction}

The computation of dynamical correlation functions at finite temperature is one of the important open problems of modern quantum many-body physics. These functions are not only crucial to test models against different spectroscopic techniques, but are also critical to the development of fast machine learning tools to accelerate and enhance understanding of problems at the forefront of condensed matter physics. For instance, the inelastic neutron scattering (INS) cross-section of quantum magnets is proportional to the dynamical spin structure factor, $S(\textbf{Q},E)$. A full calculation of this dynamical correlation function is complex and necessitates accounting for the short-range entanglement of the quantum mechanical states that describe most quantum magnets.  To surmount this challenge, semi-classical approximations such as Landau-Lifshitz dynamics (LLD) have been extensively adopted and applied~\cite{Tsai2008,Conlon2009,Samarakoon18,Samarakoon20,Lin2013,ZhangLLD2019}.

Landau-Lifshitz dynamics was originally introduced to describe the precession of the  magnetization in a solid~\cite{LandauLifshitz}. This dynamics can be derived as a classical limit of quantum spin systems, whose quantum mechanical state becomes a direct product of SU($2$) coherent states. The time evolution of this product state is dictated by the LLD equations.
It is known, however, that $N$-level quantum mechanical systems ($N=2S+1$ for spin systems) admit more than one classical limit~\cite{glauber1963a,glauber1963b,glauber1963c,sudarshan1963equivalence,perelomov1972,Gilmore1972391,Yaffe,Gnutzmann_1998}. As was pointed out in a recent work~\cite{Zhang2021}, there are large classes of low-entangled quantum magnets, such as materials with strong single-ion anisotropy, for which it is necessary to use a generalized spin dynamics (GSD) which accounts for non-dipolar components of spin states (non-zero expectation values of other multipoles). This GSD is also necessary to describe magnets with significant biquadratic interactions~\cite{muniz2014generalized,Remund22} and those comprising weakly-coupled entangled units, such as dimers~\cite{Jaime04,nawa2019}, trimers~\cite{Qiu05} and tetramers~\cite{Okamoto13,park2016,rau2016,choi2014}.
The hypothesis that the present work seeks to test is that these systems are better described by direct products of SU($N$) coherent states at any temperature.
This implies that the traditional LLD must be extended to encompass these more general cases~\cite{Zhang2021}.

\begin{figure}[t]
\centering
\includegraphics[width=0.5\linewidth]{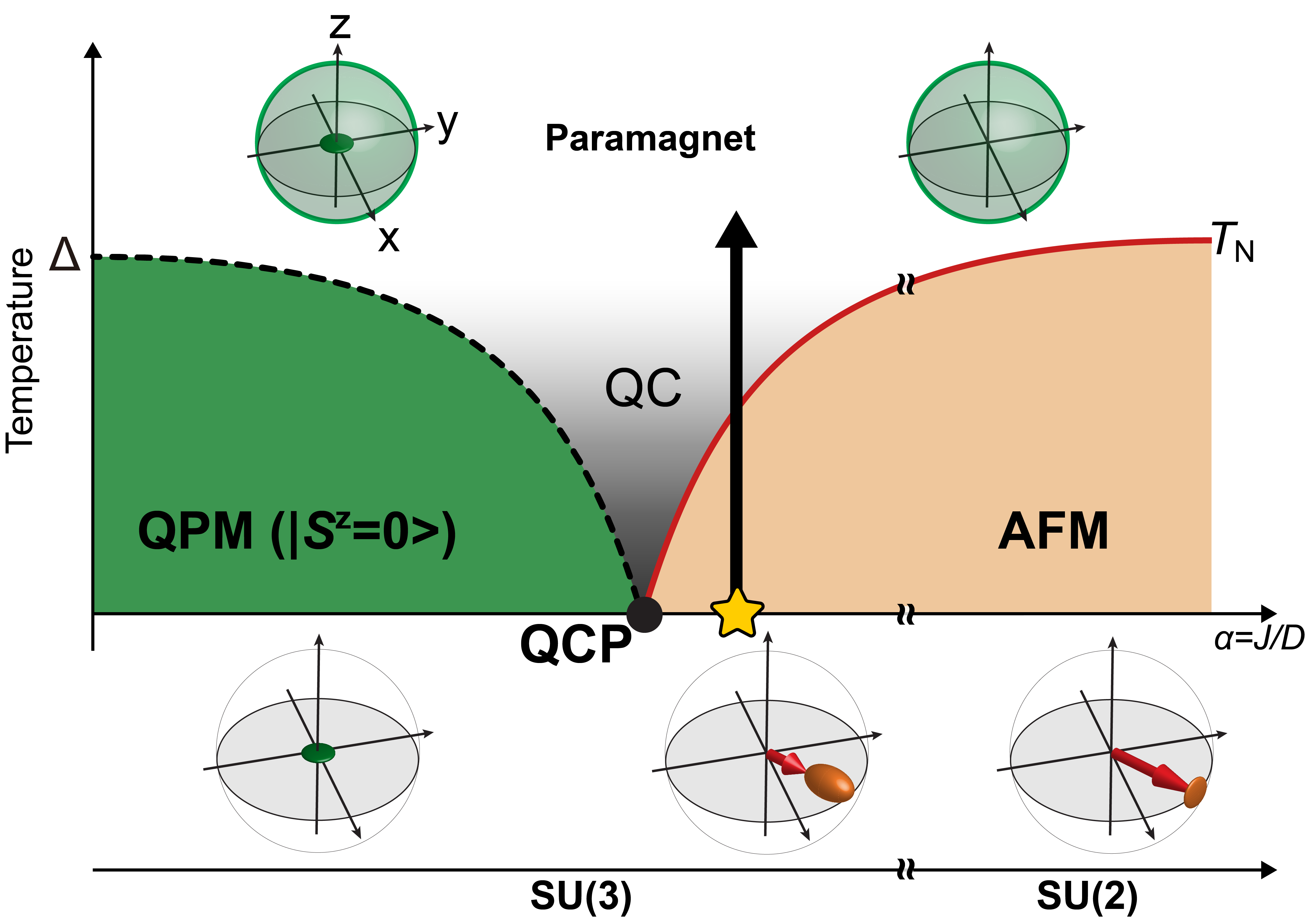} 
\caption{\label{fig:diagram} 
\textbf{Schematic phase diagram of a weakly coupled $S$=1 easy-plane antiferromagnet.}
A quantum phase transition occurs between the QPM and AFM phases as a function of $\alpha=J/D$. The Bloch spheres depict the mean-field spin state: the arrow and ellipsoid indicate the dipole moment and its fluctuations. As $\alpha \rightarrow$ 0, the spin is described by a three-level SU($3$) representation with $|S^{z}=0,\pm 1\rangle$. The $|S^{z}=0\rangle$ level of the QPM (no net dipole moment) is represented by a small disk-shape. As the temperature increases, the $|S^{z}=\pm1\rangle$ levels (green sphere) become occupied and the system crosses-over into an SU($3$) classical paramagnet. Near the QCP, the easy-plane AFM ordering is described by a mean-field product of SU($3$) coherent states with small dipole moment and large quadrupolar component (ellipsoid). Easy-plane antiferromagnet \bfso{} ($T_\text{N}=5.2$ K) is located near the QCP (yellow star). For $\alpha \gg 1$, the system is better approximated by SU($2$) coherent states.
}
\end{figure}

The main goal of this work is to test the aforementioned hypothesis by modelling the INS cross-section of the effective $S=1$ quasi-2D easy-plane antiferromagnet (AFM), \bfso{}~\cite{Do2020,Jang2021}. In \bfso{}, a significant single-ion anisotropy ($D\sim1.42$ meV) induced by the large tetragonal distortion of FeO$_4$ tetrahedron in conjunction with spin-orbit coupling of Fe$^{2+}$($3d^6$) results in an effective low-energy three-level manifold generated by the spin states, $|S^{z}=0,\pm 1\rangle$. The competition with a relatively weak Heisenberg exchange interaction, places the ground state of this material ($\alpha$=$J/D$ $\sim0.187$) near the quantum critical point at $\alpha_{c} \sim0.158$ that separates easy-planar AFM order from a quantum paramagnetic (QPM) phase (see Fig.~\ref{fig:diagram}). Since the AFM to QPM transition is driven by an enhancement of the local quadrupolar moment at the expense of the magnitude of the local dipolar moment, a proper classical description must allow for the coexistence of local dipolar and quadrupolar fluctuations, leading to transverse and longitudinal collective modes~\cite{Do2020}. SU($3$) coherent states fulfill this condition because an SU($3$) spin has $8=3+5$ components that include the three components of the dipole moment and the five components of the quadupolar moment (trace-less symmetric tensor)~\cite{Batista02,Batista04,muniz2014generalized,Bharath18,Zhang2021}. 

In this article, we use INS to explore the temperature dependence of the $S(\textbf{Q},E)$ of \bfso{}. The spin excitation spectra exhibit significant transfer of spectral weight across the N\'{e}el temperature ($T_\text{N}=5.2$ K) with a continuous evolution from well-defined SU($3$) spin-wave modes for $T < T_\text{N}$ to a diffusive resonant excitation for $T > T_\text{N}$, indicating temperature dependent SU($3$) spin dynamics. To account for this behavior, we generalize LLD to SU(3) spins. This GSD combined with Monte Carlo simulation for temperature dependent SU($3$) classical spin states~\cite{Zhang2021} describes the observed temperature dependence of $S(\textbf{Q},E)$ both above and below $T_\text{N}$. This result verifies the main hypothesis of this work, namely that many anisotropic magnets such as \bfso{} can be described by direct products of SU($N$) coherent states at any temperature.  

\section*{Results}

\begin{figure}[t]
\centering
\includegraphics[width=\linewidth]{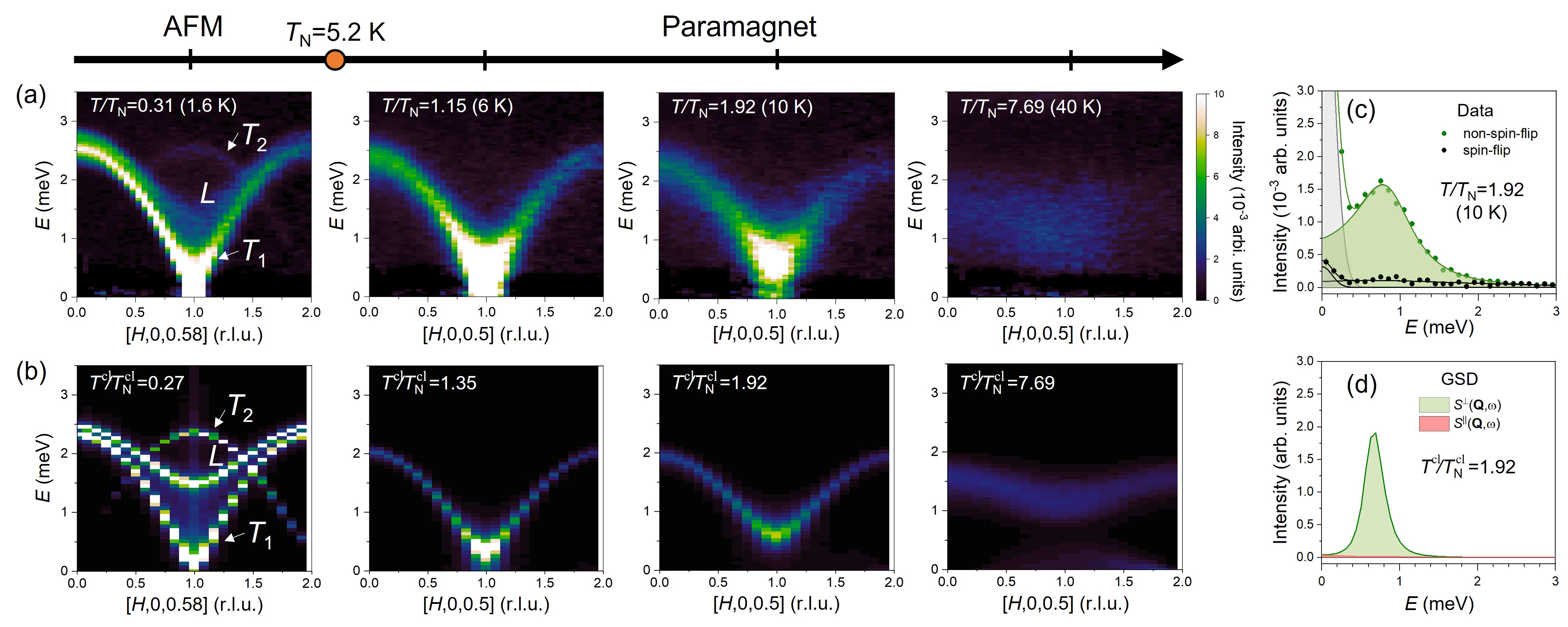}
\caption{\label{fig:T_spectra} 
\textbf{Temperature dependent INS spectra of \bfso{}.} (a) Contour maps of the INS data as a function of energy and momentum transfers along $[H,0,L]$ ($L=0.5,0.58$) measured at $T=1.6$ K, $6$ K, $10$ K, and $40$ K, corresponding to the scaled temperatures $T/T_\text{N}=0.31, 1.15, 1.92,$ and $7.69$ where $T_\text{N}=5.2$~K. Labels $T_\text{1}$, $T_\text{2}$, and $L$ indicate the acoustic and optical transverse modes, and the longitudinal mode in the spectra, respectively. (b) Resolution convoluted INS intensities calculated by the GSD method.  As described in more detail in the main text, the calculations are performed for the temperature ratio, $T^{\text{cl}}/T^{\text{cl}}_\text{N}$, indicated in each panel. 
(c) Polarized neutron scattering at \textbf{Q}$=(1,0,0)$ measured at $10$ K ($T>T_\text{N}$). The non-spin-flip (spin-flip) channel for the neutron spin polarization ($\parallel(0,1,0)$) response corresponds to the $S^{xx}+S^{yy}$ ($S^{zz}$) component of $I$($\textbf{Q}$,\textit{E}). The green-shaded region indicates the magnetic INS response extracted by fitting a DLDHO function to data (see text). (d) The green (red) shaded region indicates the $S^{xx}+S^{yy}$ ($S^{zz}$) component of $I$($\textbf{Q}$,\textit{E}) calculated by GSD.
}
\end{figure}

Figure~\ref{fig:T_spectra}(a) shows the temperature dependence of the unpolarized neutron cross-section $I$(\textbf{Q},\textit{E}) of \bfso{}, which is proportional to $S(\textbf{Q},E)$. Below $T_\text{N}$, the spectrum exhibits sharp spin-waves corresponding to acoustic ($T_1$) and optical ($T_2$) transverse modes. The longitudinal ($L$) mode is observed as a broad continuum above the $T_1$-mode throughout the entire Brillouin zone (BZ) due to decay into a pair of transverse modes as described in Ref.~[\citenum{Do2020}]. 
While the sharp spin-waves disappear above $T_\text{N}$, a broad dispersion with a finite gap emerges at the magnetic zone center (ZC), $\textbf{Q}_{m}=(1,0,0.5)$. With increasing temperature, the gap size increases and the bandwidth becomes narrower.

To understand the diffusive spectra above $T_\text{N}$, we performed a polarized neutron scattering experiment. Fig.~\ref{fig:T_spectra}~(c) shows the neutron spin polarization dependence of $I(E)$ at \textbf{Q}$=(1,0,0)$. The spin-flip and nonspin-flip scattering cross-sections are coupled to the sample magnetization and the wave-vector, allowing us to extract the directional dependence of $S$(\textbf{Q},\textit{E}). The neutron spin was polarized along $[0,1,0]$, which provides separate in-plane ($S^{\perp}=S^{xx}(\textbf{Q},E)+S^{yy}(\textbf{Q},E$)) and out-of plane ($S^{\parallel}=S^{zz}(\textbf{Q},E$)) components of $S(\textbf{Q},E)$, for nonspin-flip and spin-flip channels, respectively. The nonspin-flip channel is by far the most intense, indicating the diffusive spectra at 10 K mainly comes from the in-plane components of $S$(\textbf{Q},\textit{E}).

To account for the measured spectra at finite temperatures, we performed GSD calculations for SU($3$) spins in the three-level $|S^{z}=0,\pm 1\rangle$ representations. The low-energy effective Hamiltonian for \bfso{} is $\mathcal{H}=\sum_{\bm{r},\bm{\delta}} S^{\mu}\mathcal{J}_{\bm{\delta}}^{\mu \nu}S^{\nu}+D\sum_{\bm{r}} (S_{\bm{r}}^{z})^2$~[\citenum{Do2020}], with the convention of summation over repeated indices $\mu, \nu =\{x, y, z\}$ and
$\bm{\delta}$ runs over the neighboring bonds with finite exchange interaction.
This Hamiltonian can be recast in terms of SU($3$) generators $\hat{O}_{\bm{r}}^{\eta}$~\cite{Zhang2021}
\begin{equation}
    \mathcal{\hat{H}} = \frac{1}{2}\sum_{\bm{r},\bm{\delta}} \hat{O}_{\bm{r}}^{\eta}\mathcal{J}_{\bm{\delta}}^{\eta \gamma}\hat{O}_{\bm{r}+\bm{\delta}}^{\gamma}+\frac{D}{\sqrt{3}}\sum_{\bm{r}}\left(\hat{O}_{\bm{r}}^{8} + \frac{2}{\sqrt{3}}\right),
    \label{eq:hamiltonian}
\end{equation}
where the exchange tensor $\mathcal{J}_{\bm{\delta}}^{\eta \gamma} = \delta_{\eta \gamma} J_{\bm{\delta}} (\delta_{\eta 1}+ \delta_{\eta 2} + \delta_{\eta 3} \Delta_{\bm{\delta}})$ with $J_{\bm{\delta}} = J(J')$, $\Delta_{\bm{\delta}}=\Delta(\Delta')$ for nearest-neighbor intralayer (interlayer) bonds and $ 1\leq \eta, \gamma \leq 8$. The generators $\hat{O}_{\bm{r}}^{1-3}$ correspond to \emph{dipolar} operators $(S^x_{\bm{r}},S^y_{\bm{r}},S^z_{\bm{r}} )$ and $\hat{O}^{4-8}$ are the \emph{quadrupolar} operators (bilinear traceless forms of the dipolar operators). See Ref.~[\citenum{Zhang2021}] for the matrix representations. We note that the single-ion anisotropy becomes an external (quadrupolar) field that is linearly coupled to the SU($3$) spins.

After taking the classical limit of Eq.~\eqref{eq:hamiltonian} using SU($3$) coherent states~\cite{Zhang2021}, $\hat{O}_{\bm{r}}^{\eta}  \to o^{\eta}_{\bm{r}} \equiv \langle Z_{\bm{r}} | \hat{O}_{\bm{r}}^{\eta} | Z_{\bm{r}} \rangle $, we obtain the  classical equation of motion (EOM) of the SU($3$) spins
\begin{equation}
    \frac{d o_{\bm{r}}^\eta}{d t} 
    = \sum_{\bm{\delta}}  \mathcal{J}_{\bm{\delta}}^{\gamma \alpha} f_{\eta\gamma \lambda} o_{\bm{r}}^{\lambda} o_{\bm{r}+\bm{\delta}}^{\alpha} 
    + \frac{D}{\sqrt{3}}  f_{\eta 8 \lambda} o_{\bm{r}}^{\lambda},
\label{eq:su3LL}
\end{equation}
where $f_{\eta\gamma \lambda}$ are the SU($3$) structure constants: $[\hat{O}_{\bm{r}}^{\eta}, \hat{O}_{\bm{r}}^{\gamma}]= i f_{\eta\gamma \lambda} \hat{O}_{\bm{r}}^{\lambda}$.  To compute $S$(\textbf{Q},\textit{E}) at finite temperature, the initial conditions of Eq.~\eqref{eq:su3LL} are sampled with the standard Metropolis-Hastings Monte Carlo (MC) algorithm from the $\mathrm{CP}^2$ manifold (classical phase space) of SU($3$) coherent states (see Supplementary Information (SI)\cite{supple} for detailed information). The numerical integration methods for the classical EOM~\eqref{eq:su3LL} are explained in the SI\cite{supple} and Ref. ~[\citenum{Dahlbom2022}].
The INS intensity $I$(\textbf{Q},\textit{E}) is obtained from the Fourier transform of the classical dipolar operators $o_{\bm{r}}^{\mu}(t)\ \mu=1,2,3$~\cite{supple}.
For the calculation, we used $J=0.266$~meV, $J'=0.1J'$, and $D=1.42$~meV from Ref.~[\citenum{Do2020}] and finite lattices consisting of $24 \times 24 \times 12$  sites.  Additionally, for low values of the spin $S$, the N\'{e}el temperature of the classical spin Hamiltonian is significantly lower than the N\'{e}el temperature of the quantum mechanical Hamiltonian because quantum fluctuations further increase the energy gain of the ordered state relative to a disordered state. This is a well-known fact for the traditional classical limit based on SU($2$) coherent states which remains true for the more general case that we are considering here.  Hence the classical approximation used here underestimates the value of N\'{e}el temperature, $T^{\rm cl}_\text{N}=1.38$~K, compared to experimental value $T_\text{N}$ by a factor of $\sim 3.75$.  Therefore in Fig.~\ref{fig:T_spectra}(a) and (b) we compare the measured and calculated spectra at the same values of  $T/T_\text{N}$ and $T^{\rm cl}/T^{\rm cl}_\text{N}$, respectively.  

Below $T_\text{N}$, the calculated spectrum  exhibits $T_1$-, $T_2$- and $L$-modes (see Fig. \ref{fig:T_spectra}), where the calculated intensities by the GSD are multiplied by the classical to quantum correspondence factor for a harmonic oscillator $\beta^{\text{cl}} E /(1-e^{-\beta^{\text{cl}} E})$ with $\beta^{\text{cl}} = 1/k_\text{B} T^{\text{cl}}$~[\citenum{bohm2012}]. 
Since the GSD calculation at low-temperatures coincides with the generalized \emph{linear} spin-wave calculation~\cite{Do2020}, the decay and renormalization of the $T_2$ and $L$-modes observed at $1.6$ K (Fig.~\ref{fig:T_spectra}(a)) are not captured by this classical approximation. Capturing these feature requires the nonlinear approach described in Ref.~[\citenum{Do2020}]. Above $T_\text{N}$, the GSD calculation reproduces the gapped nature of the spectrum representing a resonant excitation between $|S^{z}=0\rangle$ and $|\pm 1\rangle$ states with a finite dispersion due to the exchange interaction. 
In the classical description, this diffusive mode originates from the combined effect of the ``external SU($3$) field'' $D$ that induces a precession of each SU($3$) moment with frequency $D/\hbar$ (center of the peak) and the random molecular field due to the exchange interaction with the fluctuating neighboring moments that determines the width of the peak. When $J \ll T$, the spectrum thus becomes a dispersion-less broad peak centered around an energy ($\Delta_\text{para}) \simeq D$. 
The computed spectra reproduces the main characteristics of the observed dispersions and bandwidth. Since the $|S^{z}=0\rangle$ and $|\pm 1\rangle$ states are connected by the components that are transverse to the $z$-axis, $S^{\pm}$=$S^{x} \pm iS^{y}$, the corresponding intensity of  $S$(\textbf{Q},\textit{E}) should appear in the channel $S^{\perp}$=$S^{xx}(\textbf{Q},E)$+$S^{yy}(\textbf{Q},E)$ (see Fig.~\ref{fig:T_spectra}(d)), which is qualitatively in good agreement with the polarization dependence of the measured $S(\textbf{Q},E)$.  

\begin{figure*}[t]
\centering
\includegraphics[width=\linewidth]{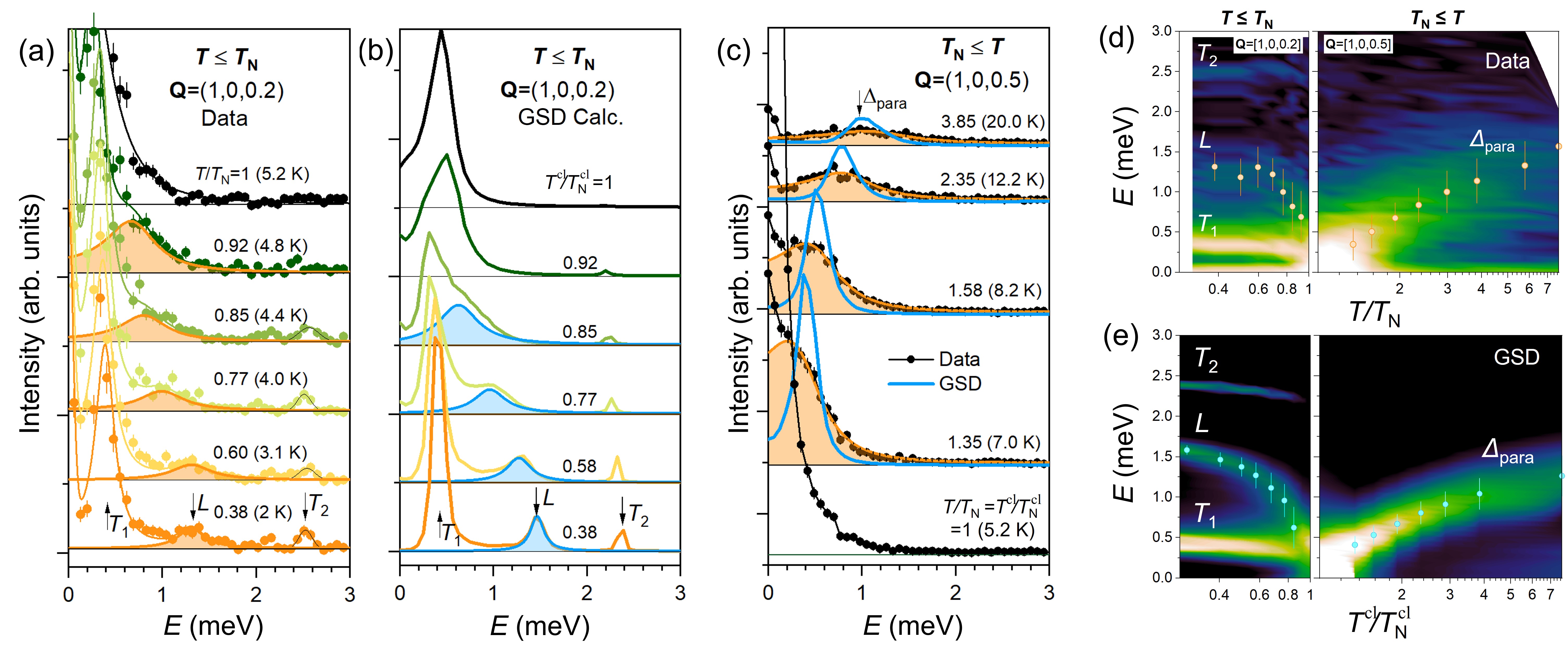}
\caption{\label{fig:lincut} 
\textbf{Detailed constant momentum scans of INS spectra.} (a) Measured constant momentum scans at \textbf{Q}$=(1,0,0.2)$ near the ZC with temperatures below $T_\text{N}$ and (b) corresponding calculated spectra by  GSD method. The transverse and longitudinal modes are indicated with labels `$T_1$', `$T_2$', and `$L$'. The spectral weight for the $L$-mode was fitted with a DLDHO function, and the results are indicated by the orange (blue) shaded regions for experimental (calculated) spectra.  The calculated intensities by the GSD in (b). 
(c) Comparison of the measured and calculated momentum scans at the ZC above $T_\text{N}$. As in (a) and (b), the spectral weights were determined by fitting with a DLDHO function and the results are indicated by the shaded regions. The extracted spectral shapes of the resonant modes are compared with the GSD calculations shown by the blue solid line. (d)(e) Comparison of contour plots of the constant momentum scans between INS data and GSD calculation across $T_\text{N}$ and $T_\text{N}^\text{cl}$. The position ($\Delta$) and line-width ($\Gamma$) of the $L$-mode and resonant excitation were quantified by the DLDHO function, and are exhibited as points with error-bars, respectively.
}
\end{figure*}

The detailed spectral change across $T_\text{N}$ is shown in Fig.~\ref{fig:lincut} which compares the measured and calculated constant momentum scans at the ZC with varying temperature. 
For $T<T_\text{N}$, we consider a  wave vector \textbf{Q}$=(1,0,0.2)$ that is close but not exactly equal to the ZC $\textbf{Q}=(1,0,0.5)$ in order to avoid the large tail of the magnetic Bragg reflection as well as the technical challenges associated with calculating the spectrum at the ZC. In this case, the $T_1$ (Goldstone) mode becomes visible because of its finite energy at \textbf{Q}$=(1,0,0.2)$ due to the non-zero $[0,0,L]$-dispersion produced by the small inter-layer coupling $J_\text{inter}$. As a result, the three $T_1$, $T_2$, $L$ modes are observed in the spectrum (see Fig.~\ref{fig:lincut}(a)). 
While the $T_1$ and $T_2$ transverse modes remain nearly unchanged with increasing temperature, the energy of the $L$-mode decreases and the mode becomes broader and indistinguishable from the quasielastic scattering near $T_\text{N}$. Above $T_\text{N}$, the quasielastic scattering continuously evolves into a broad peak centered at finite energy ($\Delta_\text{para}$), whose energy increases gradually with the temperature (see Fig.~\ref{fig:lincut}(c)).
To extract the spectral weight of the resonant excitation above $T_\text{N}$ and the $L$-mode below $T_\text{N}$, the data were fitted with a double Lorentzian function associated with a damped harmonic-oscillator (DLDHO),  
\begin{align}
S(E)=A(n(E)+1)\left[\frac{\Gamma}{(E-\Delta)^{2}+\Gamma^{2}}-\frac{\Gamma}{(E+\Delta)^{2}+\Gamma^{2}} \right], 
\label{dldho}
\end{align}
that provides a simplified description of the contribution of an over-damped mode~\cite{Hong2008,Hong2017,supple}. The $n(E)+1$ is the Bose factor, and $A$, $\Delta$, and $\Gamma$ indicate the intensity, energy, and line-width of the peak, respectively. The extracted spectral weights and parameters of the $L$-mode and the resonant excitations are indicated by the shaded regions in Fig.~\ref{fig:lincut}(a)-(c), and summarized in Fig.~\ref{fig:lincut}(d)(e), respectively.

\begin{figure}[t]
\centering
\includegraphics[width=0.5\linewidth]{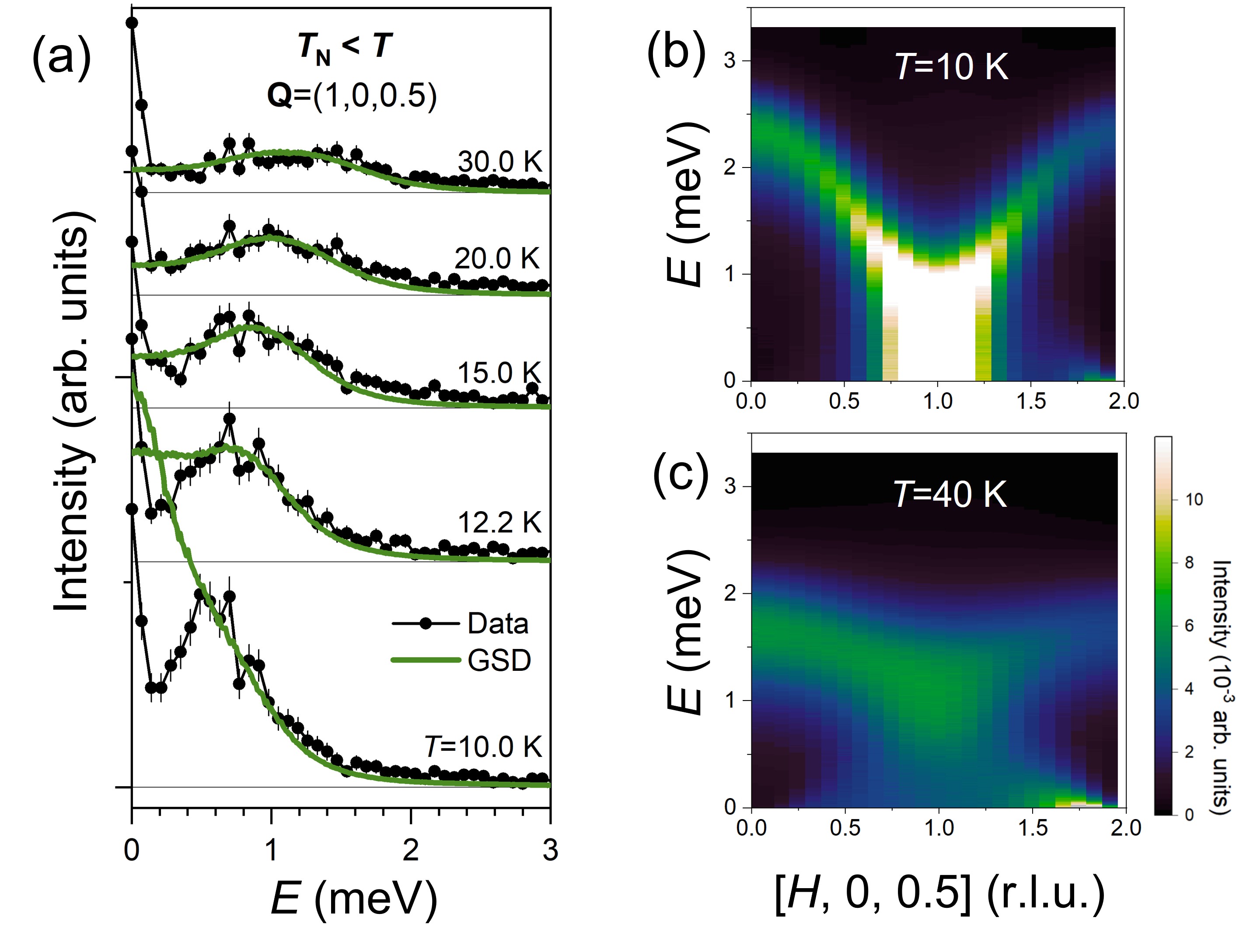}
\caption{\label{fig:scaling_factor} 
\textbf{Simulated INS spectra with the renormalized SU($3$) spins.} (a) Comparison of the measured constant momentum scans (black dots) at the ZC above $T_\text{N}$ with the results obtained from GSD simulations (green solid lines) with renormalized SU($3$) spins ($o_{\bm{r}}^{\eta}  \to \kappa o^{\eta}_{\bm{r}}$ with $\kappa=2$), where the recalculated theoretical N{\'e}el temperature $T_\text{N}^\text{cl}$ for $\kappa=2$ is $T_\text{N}^\text{cl}=7.5$~K (see text). Simulated \textit{I}(\textbf{Q},\textit{E}) (b) at $T=10$ K and (c) at $T=40$ K with the renormalized SU($3$) spins. 
}
\end{figure}

Figure~\ref{fig:lincut}(b)(c) show a comparison of the GSD calculation with the INS data across $T_\text{N}$. Remarkably, the spectral weight for the $L$-mode is enhanced and shifts to low-energy with increasing temperature, which is consistent with the data. Above $T_\text{N}$, the GSD calculation gives a diffusive resonant peak-shaped spectrum centered at the energy $\Delta_\text{para}$ that approaches $D$ for $T \gg T_\text{N}$. We note that the traditional LLD based on SU($2$) coherent states cannot explain this gapped diffusive mode as well as the $L$-mode, leading to incorrect result in the high-temperature limit~\cite{Zhang2021}. 
However, we notice that the calculated spectrum underestimates the width of the mode at temperatures $T \gtrsim T_\text{N}$. This discrepancy arises from an inadequate normalization of the SU($3$) spins at $T \gtrsim T_\text{N}$~\cite{Zhang2021}, similar to the issue raised in traditional SU($2$) LLD~\cite{Huberman08}.

This discrepancy in the line-width can be removed in the high-temperature limit by applying an adequate renormalization of the SU($3$) spins $o_{\bm{r}}^{\eta}  \to \kappa o^{\eta}_{\bm{r}}$, with $\kappa=2$ in the high-temperature ($T \gg T_\text{N}$) limit, as described in Ref.~[\citenum{Zhang2021}]. This renormalization guarantees that the SU($3$) $S(\textbf{Q},E)$ satisfies the exact sum-rule in the high-$T$ limit. Properly renormalizing the spin has the additional virtue of bringing the theoretical N\'{e}el temperature to $T^{\rm cl}_\text{N}~\simeq 7.5$ K ($\kappa=2$) closer to the experimental value $T_\text{N}=5.2$~K.
Fig.~\ref{fig:scaling_factor}~(a) shows the comparison between the new calculations including the renormalization factor and the data measured at the same temperature $T^{\rm cl}= T$ (experiment). This comparison reveals a better agreement in spectral shape at $T_\text{N} \ll T$ than those without renormalization factor (see Fig.~\ref{fig:lincut}(c)). As expected, significant deviations in low-energy spectrum are observed at $T=10$ K and $12$ K because they are relatively close to the $T^{\rm cl}_\text{N}$ which deviates from the experimental $T_\text{N}$.
In other words, since the energy $\Delta_\text{para}$ of the diffusive peak goes to zero at $T_\text{N}$, its position is shifted to the lower energy relative to the measured peak at $T=10$ K. The emergence of a quasielastic peak (centered at $E=0$) in the theoretical calculation indicates proximity to the $T^{\rm cl}_\text{N}$. Figures~\ref{fig:scaling_factor}~(b) and (c) show the computed INS cross-section, \textit{I}(\textbf{Q},\textit{E}), at $T=10$ K and $40$ K along the same direction in reciprocal space that is presented in Fig.~\ref{fig:T_spectra}~(a). The $T=10$~K and $40$~K spectra with renormalized spin provide a good description of the measured spectra at these temperatures.


In summary, the GSD based on SU($3$) spin provides a good approximation of the measured INS cross-section over a broad temperature range. The most quantitative deviations are observed at very low-temperatures $T \ll T_\text{N}$ and close to $T_\text{N}$. The former case is due to the requirement of quantum correction to account for the decay of the $L$-mode~\cite{Do2020}, and the latter is due to the expected discrepancy between the experimental and re-scaled values of $T_\text{N}$ originated from the renormalization factor $\kappa=2$ to the classical SU($3$) spins. 
This renormalization factor arises from enforcing the sum-rule in the infinite $T$-limit. Similarly, the GSD of unrenormalized classical SU($3$) ($\kappa=1$) leads to the correct sum-rule in the zero temperature limit after quantizing the normal modes. Therefore, the correct scaling factor should be defined as a function $\kappa(T)$ that monotonically interpolates between the two limiting cases $\kappa(0)=1$ and  $\kappa(\infty)=2$.

The verification of the main hypothesis of this work has very important consequences for the characterization of quantum magnets. For instance, while INS is an ideally suited technique for extracting models from data, the solution of this “inverse scattering problem” requires the development of fast solvers of the direct problem (inferring the INS cross-section of a given model). A crucial advantage of the GSD demonstrated here is that the cost of the simulations increases linearly in the system size~\cite{Dahlbom2022}, making it an ideal solver for attacking the inverse scattering problem with machine-learning-based approaches~\cite{Samarakoon20}. Moreover, since the GSD can reproduce the INS data in the high-temperature regime, the method can still be applied to quantum magnets that exhibit long-range entanglement at low enough temperatures, but undergo a quantum to classical crossover above a certain temperature $T_\text{QC}$~\cite{Samarakoon18,Samarakoon18b}. Finally, we note that the SU($N$) approach described here is also relevant to the broad class of materials comprising weakly coupled antiferromagnetic magnets including dimers, trimers, or tetramers as well as magnets with strong single-ion anisotropy, where similar effects may be anticipated.

\section*{Methods}

For the INS experiments, the same single crystal of \bfso{} (mass: $2.13$~g) as was used in Ref. [\citenum{Do2020}] was aligned on an aluminum plate with an $[H,0,L]$ horizontal scattering plane. Unpolarized INS data were collected using the cold neutron triple-axis spectrometer (CTAX) at the High Flux Isotope Reactor (HFIR) and the hybrid spectrometer (HYSPEC) at the Spallation Neutron Source (SNS) located at Oak Ridge National Laboratory~\cite{hyspec}. A liquid helium cryostat was used to control temperature for both experiments. At CTAX, the initial neutron energy was selected using a PG $(002)$ monochromator, and the final neutron energy was fixed to $E_\text{f}= 3.0$ meV by a PG $(002)$ analyzer. The horizontal collimation was guide$-$open$-40'-120'$, which provides an energy resolution with full width half maximum (FWHM)$=0.1$ and $0.18$~meV for $E=0$ and $2.5$ meV, respectively. For the HYSPEC experiment, $E_\text{i}=9$ meV and a Fermi chopper frequency of $300$ Hz were used, which gives FWHM$=0.28$~meV and $0.19$~meV of energy resolution at $E=0$ and $2.5$~meV, respectively. Measurements were performed by rotating the sample from $-50^{\circ}$ to $170^{\circ}$ with $1^{\circ}$ steps. Data were integrated over $K=[-0.16,0.16]$ and $L=[L-0.1,L+0.1]$, and symmetrized over positive and negative $H$. 

We also performed polarized neutron scattering measurement as part of the HYSPEC experiment using XYZ-polarization analysis which is same configuration as the experiment in Ref. [\citenum{soda2018}]. The X-axis is defined along \textbf{Q}$=[1,0,0]$ for the scattering wave vector, the Z-axis is defined along $[0,1,0]$ perpendicular to the scattering plane, and the Y-axis is defined along the direction $[0,0,1]$ perpendicular to the X- and Z-axes. In the experiment, the neutron was polarized along the Z-direction, and the nonspin-flip and spin-flip scattering cross-sections provide $I_\text{n}(E)+I^\text{Z}_\text{mag}(E)$ and $I^\text{Y}_\text{mag}(E)$, respectively~\cite{moon1969,soda2018}. The $I_\text{n}(E)$ indicates non magnetic structure factor and $I_\text{mag}^{\alpha}(E)$, where $\alpha \in \{\text{X,Y,Z}\}$, is the component dependent magnetic structure factor. The measured spin-flip and nonspin-flip cross-sections provide distinct $I^\text{Z}_\text{mag}(E)$ and $I^\text{Y}_\text{mag}(E)$, which correspond to $S^{xx+yy}(E)$ and $S^{zz}(E)$, respectively, on the crystal axis where $x \parallel a$, $y \parallel b$, and $z \parallel c$ in the tetragonal crystal structure.
All of data sets were reduced and analyzed using the MANTID~\cite{mantid} and DAVE~\cite{dave} software packages.

\bibliography{main}

\section*{Acknowledgements}
This work was supported by the U.S. Department of Energy, Office of Science, Basic Energy Sciences, Materials Science and Engineering Division. This research used resources at the High Flux Isotope Reactor and Spallation Neutron Source, DOE Office of Science User Facilities operated by the Oak Ridge National Laboratory (ORNL). The work at Max Planck POSTECH/Korea Research Initiative was supported by Nano Scale Optomaterials and Complex Phase Materials (2016K1A4A4A01922028) and Grant No. 2020M3H4A2084418, through the National Research Foundation (NRF) funded by MSIP of Korea. The work at Rutgers University was supported by the DOE under Grant No. DOE: DE-FG02-07ER46382. D. D., K. B. and C.D.B.~acknowledge support from U.S. Department of Energy, Office of Science, Office of Basic Energy Sciences, under Award No.~DE-SC0022311.

\section*{Author contributions statement}
S.H.D., H.Z., C.D.B., and A.D.C. conceived the project.
T.H.J., S.W.C., and J.-H.P. provided single crystals. S.H.D., T.J.W., T.H., V.O.G., and A.D.C. performed INS experiments. S.H.D., and A.D.C. analyzed the neutron data. H.Z., D.A.D, K.B., and C.D.B. constructed theoretical model and calculations. S.-H.D., H.Z., C.D.B., and A.D.C. wrote the manuscript with
input from all authors.

\section*{Additional information}
\textbf{Supplementary information} is available for this paper.


\clearpage
\addtolength{\oddsidemargin}{-0.75in}
\addtolength{\evensidemargin}{-0.75in}
\addtolength{\topmargin}{-0.725in}

\newcommand{\addpage}[1] {
\begin{figure*}
  \includegraphics[width=8.5in,page=#1]{Supplementary_information.pdf}
\end{figure*}
}

\addpage{1}
\addpage{2}
\addpage{3}

\end{document}